\documentclass[12pt,a4p]{article}
\usepackage{epsfig}

\newlength{\picwi}
 \newcommand{\tcaption}[1]{
        \refstepcounter{table}
        \setbox\@tempboxa = \hbox{\footnotesize \bf Table~\thetable. #1}
        \ifdim \wd\@tempboxa > 6in
           {\begin{center}
        \parbox{6in}{\footnotesize\baselineskip=12pt \bf Table~\thetable. #1}
            \end{center}}
        \else
             {\begin{center}
             {\footnotesize \bf Table~\thetable. #1}
              \end{center}}
        \fi}

\def\etau{$\rm e\tau $ }
\def\mutau{$\rm \mu\tau $ }
\def\emu{$\rm e\mu $ }
\def\ee{$\rm e^+e^- $ }
\def\mumu{$\rm \mu^+\mu^- $ }
\def\bm{\mathbf}
\def\tautau{ $\rm \tau^+\tau^- $ }

\def\ga{\gamma }
\def\give{\rightarrow}

\begin{document}
\begin{titlepage}
\begin{center}{\large   EUROPEAN ORGANIZATION FOR NUCLEAR RESEARCH
}\end{center}\bigskip
\begin{flushright}
        CERN-EP-2001-061   \\ 02 August 2001
\end{flushright}
\bigskip\bigskip\bigskip\bigskip\bigskip
\begin{center}{\huge\bf  
Search for Lepton Flavour Violation in 
$\mathbf{e^+e^-} $ Collisions at $\mathbf{ \sqrt{s}= 189\ -\ 209\ }$ GeV \\ 
}\end{center}\bigskip\bigskip
\begin{center}{\LARGE The OPAL Collaboration
}\end{center}\bigskip\bigskip
\bigskip\begin{center}{\large  Abstract}\end{center}
  We search for lepton flavour violating events (\emu, \etau and 
  $\rm \mu\tau)$ that 
  could be directly produced in \ee annihilations, using 
  the full available data sample collected with the 
  OPAL detector at centre-of-mass energies between 
  189 GeV and 209 GeV. 
  In general, the 
  Standard Model expectations describe the data well for all the
  channels and at each $\rm \sqrt{s}.$ A single \emu
  event is observed where according
  to our Monte Carlo simulations only 0.019 events are
  expected from Standard Model processes.
  We obtain the first limits on 
  the cross-sections $\rm \sigma(e^+e^-\give e\mu,\ e\tau\ and\ \mu\tau) $ 
  as a function of $\rm \sqrt{s} $ at LEP2 energies.  
\bigskip\bigskip\bigskip\bigskip
\bigskip\bigskip
\begin{center}{\large
(Submitted to Physics Letters B)
}\end{center}
\end{titlepage}
\begin{center}{\Large        The OPAL Collaboration
}\end{center}\bigskip

\begin{center}{\small
G.\thinspace Abbiendi$^{  2}$,
C.\thinspace Ainsley$^{  5}$,
P.F.\thinspace {\AA}kesson$^{  3}$,
G.\thinspace Alexander$^{ 22}$,
J.\thinspace Allison$^{ 16}$,
G.\thinspace Anagnostou$^{  1}$,
K.J.\thinspace Anderson$^{  9}$,
S.\thinspace Arcelli$^{ 17}$,
S.\thinspace Asai$^{ 23}$,
D.\thinspace Axen$^{ 27}$
% ...
}\end{center}\bigskip
\bigskip
{\small
$^{  1}$School of Physics and Astronomy, University of Birmingham,
Birmingham B15 2TT, UK
\newline
$^{  2}$Dipartimento di Fisica dell' Universit\`a di Bologna and INFN,
I-40126 Bologna, Italy
\newline
$^{  3}$Physikalisches Institut, Universit\"at Bonn,
D-53115 Bonn, Germany
\newline
% ...
$^{  5}$Cavendish Laboratory, Cambridge CB3 0HE, UK
\newline
% ...
$^{  9}$Enrico Fermi Institute and Department of Physics,
University of Chicago, Chicago IL 60637, USA
\newline
% ...
$^{ 16}$Department of Physics, Schuster Laboratory, The University,
Manchester M13 9PL, UK
\newline
$^{ 17}$Department of Physics, University of Maryland,
College Park, MD 20742, USA
\newline
% ...
$^{ 22}$Department of Physics and Astronomy, Tel Aviv University,
Tel Aviv 69978, Israel
\newline
$^{ 23}$International Centre for Elementary Particle Physics and
Department of Physics, University of Tokyo, Tokyo 113-0033, and
Kobe University, Kobe 657-8501, Japan
\newline
% ...
$^{ 27}$University of British Columbia, Department of Physics,
Vancouver BC V6T 1Z1, Canada
\newline
% ...
$^{  e}$ and Heisenberg Fellow
\newline
% ...
Tel Aviv 69978, Israel.
        }
\section{ Introduction} 
   Within the minimal Standard Model (SM), the fermion mass matrices  
   and the mechanism of electroweak symmetry breaking remain unexplained. 
   The conservation 
   of lepton number separately for each generation has no strong 
   theoretical basis. In addition, recent data~\cite{kamioka} present
   evidence for neutrino oscillations which necessarily violate
   lepton-flavour symmetry.
   Beyond the SM, lepton flavour violation (LFV) can occur in 
   many supersymmetric (SUSY) extensions. An example is the 
   SO(10) SUSY GUT model~\cite{so10} where both the left- and
   right-handed supersymmetric lepton partners induce LFV, but none 
   of the existing models predicts a measurable effect at LEP2 energies. 
   Experimentally, no evidence for direct LFV has been   
   reported so far. Upper bounds for muon decays are
    $\rm BR(\mu^- \give e^-\ga)< 1.2 \times 10^{-11} $ and   
   $\rm BR(\mu^+\give e^+e^+e^-)< 10^{-12}$~\cite{PDG}.  
   Searches for neutrinoless $\rm \tau $ decays~\cite{cleo-argus}
   such as 
   $\rm \tau^+\give e^+e^+e^-$ and 
   $\rm \tau^+\give \mu^+\mu^+\mu^- $  yield upper limits
   $\rm BR(Z\give e\tau)< 5.4\times 10^{-5} $ and 
   $\rm BR(Z\give \mu\tau)<7.1\times 10^{-5} $ at $\rm 90\% $ 
   confidence level (CL). Direct searches in \ee annihilations at
   the $\rm Z $ peak 
   performed by the  LEP experiments~\cite{LEP} yielded 
   $\rm 95\% $ CL limits 
    $\rm BR(Z \give e\mu,\ e\tau,\ 
   \mu\tau)< \cal{O}$ $\rm (1) \times 10^{-5}.$
   In this paper we search for the \emu, \etau and \mutau final states in 
   \ee collisions at centre-of-mass energies between 189
    GeV and 209 GeV. \par   
\section{Data sample  and event simulation}
  The OPAL detector is described in detail in~\cite{opaldet}. 
  In the present analysis, the silicon micro vertex detector, the 
  central tracking chambers, the
  electromagnetic calorimeter, the hadron calorimeter and the 
  muon chambers were required to be fully operational.
  The full data sample 
  collected since 1998 at $\rm \sqrt{s}=189 $ GeV and above
  is analysed. The corresponding collected integrated luminosities
  are shown in Table 1.\par
  Track reconstruction is performed by combining the information from the 
  silicon micro vertex detector, the vertex drift chamber,  the 
  large volume jet drift chamber and an outer layer of drift chambers
  for the measurement 
  of the $z$ coordinate\footnote{The OPAL coordinate system 
  is defined so that the $z$ axis is in the
  direction of the electron beam, the $x$ axis 
  points towards the centre of the LEP ring, and  
  $\theta$ and $\phi$
  are the polar and azimuthal angles, defined relative to the
  $+z$- and $+x$-axes, respectively. 
  In cylindrical polar coordinates, the radial coordinate is denoted
  $r$.}
  The OPAL electromagnetic calorimeter (ECAL) consists of a barrel 
  part covering the region $|\cos{\theta}| < 0.82 $ and two endcaps
  covering the region $0.82<|\cos{\theta}|<0.98$. A set of forward detectors
  provide complementary coverage for $\rm \theta> 25\ mrad.$\par
  Detector acceptance and reconstruction efficiencies for 
  processes under study 
  are evaluated with two different methods based on 
  Monte Carlo simulations. They 
  are cross-checked with a third method that uses 
  data only, where possible.
  The first method consists of generating   
  \emu, \etau and \mutau signal events
  with the EXOTIC~\cite{exotic} generator using isotropic angular 
  distributions. The second method uses Standard Model lepton pair final state
  events.
  The lepton pairs were simulated using kk2f~\cite{kk2f} 
  and KORALZ~\cite{koralz} for 
  $\rm \tau\tau(\ga) $  and $\rm \mu\mu(\ga) $ and using BHWIDE~~\cite{bhwide}
  and TEEGG~\cite{teegg} 
  for $\rm ee(\ga)$. 
  Events passing the preselection are mixed to obtain
         \emu, \etau and \mutau topologies.  The mixing assumes
         a uniform detector response in phi, and events are
         mixed only if the polar angles of their thrust axes
         are within 2 degrees of each other.  The mixing is 
         performed by rotating the momentum components in the 
         plane transverse to the beam axis of a lepton from one 
         event to match the lepton replaced in the mixed event.
  For each channel,   
  the efficiency is estimated in bins of $\rm \cos{\theta} $ and then 
  averaged giving the same weight to each bin, in order to obtain
  a value that corresponds to a uniform angular distribution of the 
  particles in the final state. The third method is the same as the second 
  method but it uses a high purity sample of lepton pair events from the
  real data.
  The efficiencies after the final event selection (described in 
  the next section) are almost independent of the 
  centre-of-mass energy. They are 
  shown in Table 1, after averaging for 
  different  centre-of-mass energies, together with  
  their statistical and systematic uncertainties added in quadrature.\par
  The SM background contributions are evaluated with large 
  samples of events processed through a full simulation of the 
  OPAL detector~\cite{opalsim} and analysed using the same reconstruction 
  and selection programs as applied to the data. The SM sample comprises 
  lepton pair final states with initial and final state radiation,
  $\rm q\bar{q}(\ga)$ events generated with PYTHIA~\cite{phytia}, a full set 
  of four-fermion final states generated with grc4f~\cite{grc4f}
  and KORALW~\cite{koralw} 
  and gamma-gamma scattering events generated using HERWIG~\cite{herwig} 
  and PHOJET~\cite{phojet}.
  SM Monte Carlo processes were generated at
  $\rm \sqrt{s} = 189\ GeV,\ 192\ GeV,\ 196 $ GeV and in steps of 2 GeV from 
  200 GeV to 208 GeV.  
  The total generated SM sample corresponds to more than 500 times
  the integrated luminosity of the recorded data for \tautau, \mumu, 
  $\rm q\bar{q}$ and 4-fermion final states and to about 50 times
  the integrated luminosity of the recorded data for \ee final states.\par
\section{Event selection}
   Events with at least two and at most 8 measured tracks 
   and no isolated photons in the electromagnetic calorimeters are 
   considered for the analysis. Isolated photons are defined as in ~\cite{ww}.
   Each track should be consistent with originating from 
   the interaction point with a measured momentum, $\rm P_{trk}, $
   exceeding 250 MeV. An energy flow algorithm~\cite{mt,ecor}
   is used to measure each track and cluster energy, and to correct for 
   possible double counting.        
   The cone jet finder~\cite{cone}, with a cone half angle of 
   $\rm 15^{\circ} $ and a minimum cone energy of $\rm 10\% $ of the  
   beam energy, $\rm E_{beam}$, is used and 2 jets are required
   in each event.
   Here, a single isolated track 
   can form a jet if it has more energy than the minimum required 
   energy in the cone. 
   To gain in the energy resolution and 
   remove potential background, namely Bhabha and gamma-gamma
   scattering, the momentum vector of each jet must satisfy 
   $\rm |\cos{\theta}|\le 0.82$.\par 
   All three search channels have in common the feature that one 
   of the event hemispheres should consist of a single 
   electron or a single muon with a measured momentum 
   close to $\rm E_{beam}.$ Events are selected  
   for further analysis if:
\begin{itemize}
\item   
   The total measured energy outside the two cones defining the jets
   is smaller than $\rm 0.10 \times E_{beam}$, the sum of the 
   two jet energies is larger than $\rm E_{beam} $ and the event
   thrust is greater than 0.95.  
   To have sensitivity to events produced at 
   effective centre-of-mass energies lower than the actual  $\rm \sqrt{s}, $
   we accept events for which 
   the total measured energy in both ends of the forward detectors is up to
   10 GeV. 
\item   
   One of the two jets has its energy greater than 
   $\rm 0.8 \times E_{beam}.$ That jet should contain a single 
   isolated track, defined as a track with 
   a momentum greater than $\rm 0.5\times E_{beam} $ located in 
   a cone with a $\rm 10^{\circ} $ half opening angle. The  $\rm 15^{\circ} $ cone should not 
   contain any other track that has a momentum in excess 
   of $\rm 0.03\times E_{beam}$.  
\end{itemize} 
   After these cuts most of the  
   hadronic final states and the $\rm \gamma\gamma $ scattering events
   are rejected and the remaining event sample 
   consists of $\rm 92\%$ wide-angle Bhabha scattering events and $\rm\simeq 8\% $ 
   \mumu and \tautau events. 
   The distributions of the visible energy (defined as the sum 
   of all measured track energies),
   the total measured energy  
   in the electromagnetic calorimeter and the $\rm \cos{\theta}$ 
   of the total momentum are shown in 
   Figure~\ref{fig:plots}.
   Here, the total momentum is defined as 
   the vector sum of all measured momenta.     
   The SM expectation describes the measured data well.\par
   A track is considered to be the electron candidate if 
   it has an associated 
   energy in the electromagnetic calorimeter, $\rm E_{ECAL},$
   within $\rm 20\% $ of $\rm E_{beam}$  
   and if the ratio $\rm E_{ECAL}/P_{trk} $ is greater than 0.7. 
   The track should also have 
   a characteristic ionisation in the tracking chambers consistent with an
   electron hypothesis. 
   A track is considered to be  
   a muon candidate if it has matching hadron calorimeter and
   muon chamber hits and if the ratio  $\rm  E_{ECAL}/ P_{trk} $ is less
   than 0.1. The  following cuts, common to all three search channels, are applied:
\begin{figure}
       \centerline{\includegraphics[scale=0.8]{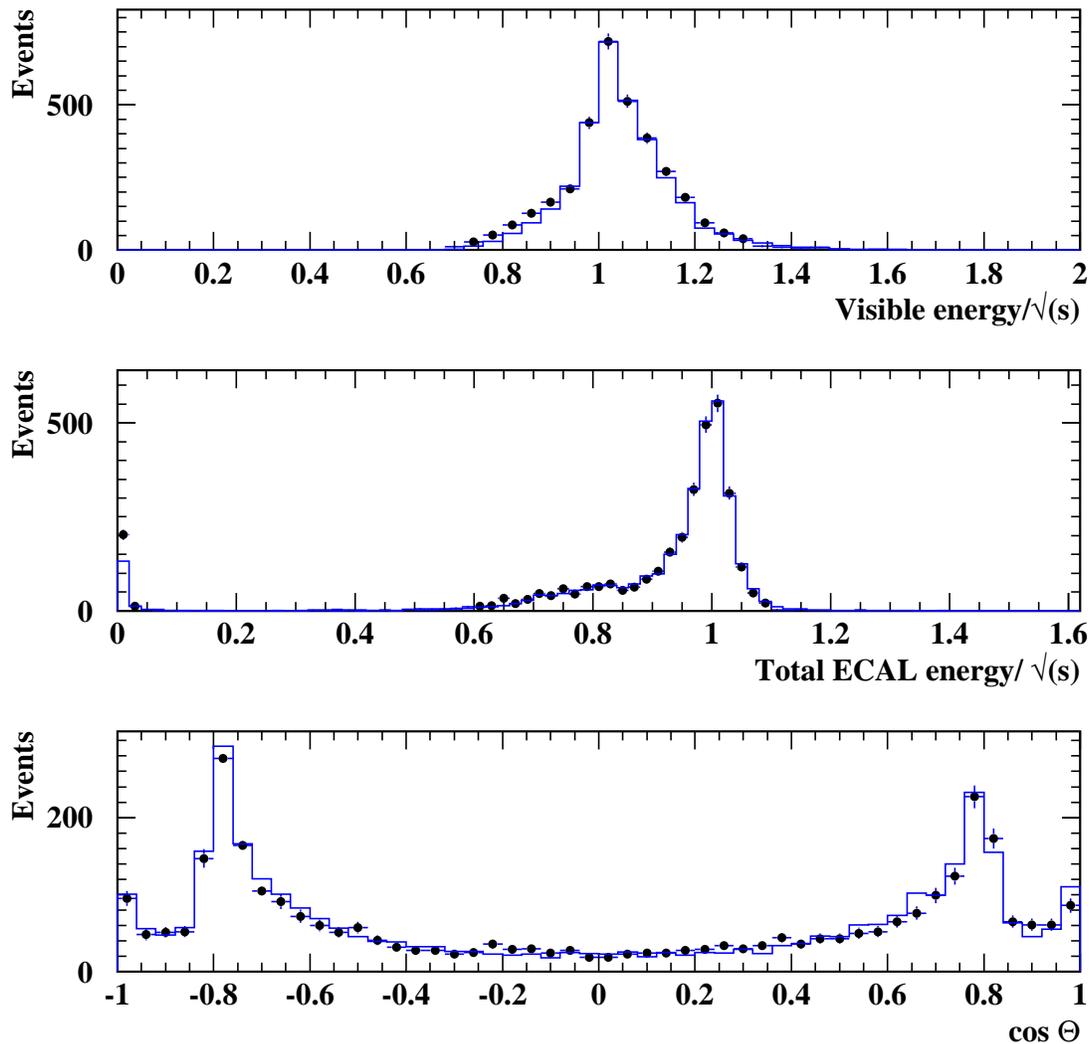}}
       \caption{ The measured visible energy, 
         the total measured energy in the electromagnetic calorimeter
         and the $\rm\mathbf{\cos{\theta}} $ of the total momentum 
         for all data (points) as 
         compared to the Standard Model Monte Carlo 
         expectation (solid lines). The small asymmetry in the total 
         momentum distribution
         is caused by the fact that the interaction
         vertex is not the geometrical center of the OPAL detector.
         They are about $\rm \simeq 1\ cm $ 
        away from each other. But this effect is
         well modelled in the Monte Carlo.  
         }   
       \label{fig:plots}
\end{figure} 
\begin{enumerate}  
\item To reject the Bhabha scattering, events are selected if
      the total measured energy in the electromagnetic 
      calorimeter is less than $\rm 1.6\times E_{beam}. $
\item Events that have a reconstructed electron candidate in the transition
      region
      between the barrel and endcaps ($\rm 0.75<|\cos{\theta}|< 0.85$) 
      are rejected, ensuring that the remaining electron
      candidates have a good energy resolution.
\item To reduce the \ee and  \mumu background further, events that have 
      two muon or two electron candidates are rejected. This reduces 
      the \etau and \mutau efficiencies by about $\rm 15\% $ each,
      rejecting events with   
      $\rm \tau\give \nu \bar{\nu} e\ or\ \nu\bar{\nu}\mu $ 
      where the final state electron or
      muon satisfies the requirement of an isolated track associated 
      with an electron or a muon. 
\item To suppress the \tautau final states, events with two isolated tracks
      are rejected if each track has a measured momentum 
      less than $\rm 0.30\times E_{beam}$. They are also rejected if 
      the opening angle between the two isolated tracks is less than 
      160 degrees.
\end{enumerate}
     The distribution of $\rm E_{ECAL}/E_{beam} $ for electron candidates
     and the distribution of
     $\rm P_{trk}/E_{beam} $ for muon candidates are shown in Figure 2.
     These distributions are obtained for the selected events
     after cut 1 above and show good agreement
     between data and SM expectations.   
     Using Monte Carlo lepton pair final states, we find that 
     the resolution on $\rm E_{ECAL} $ is 
     $\rm 3.5 \% $ for electrons  and the resolution 
     on  $\rm P_{trk} $ is $\rm 11\% $ 
     for muons.\par
\begin{figure}
       \centerline{\includegraphics[scale=0.8]{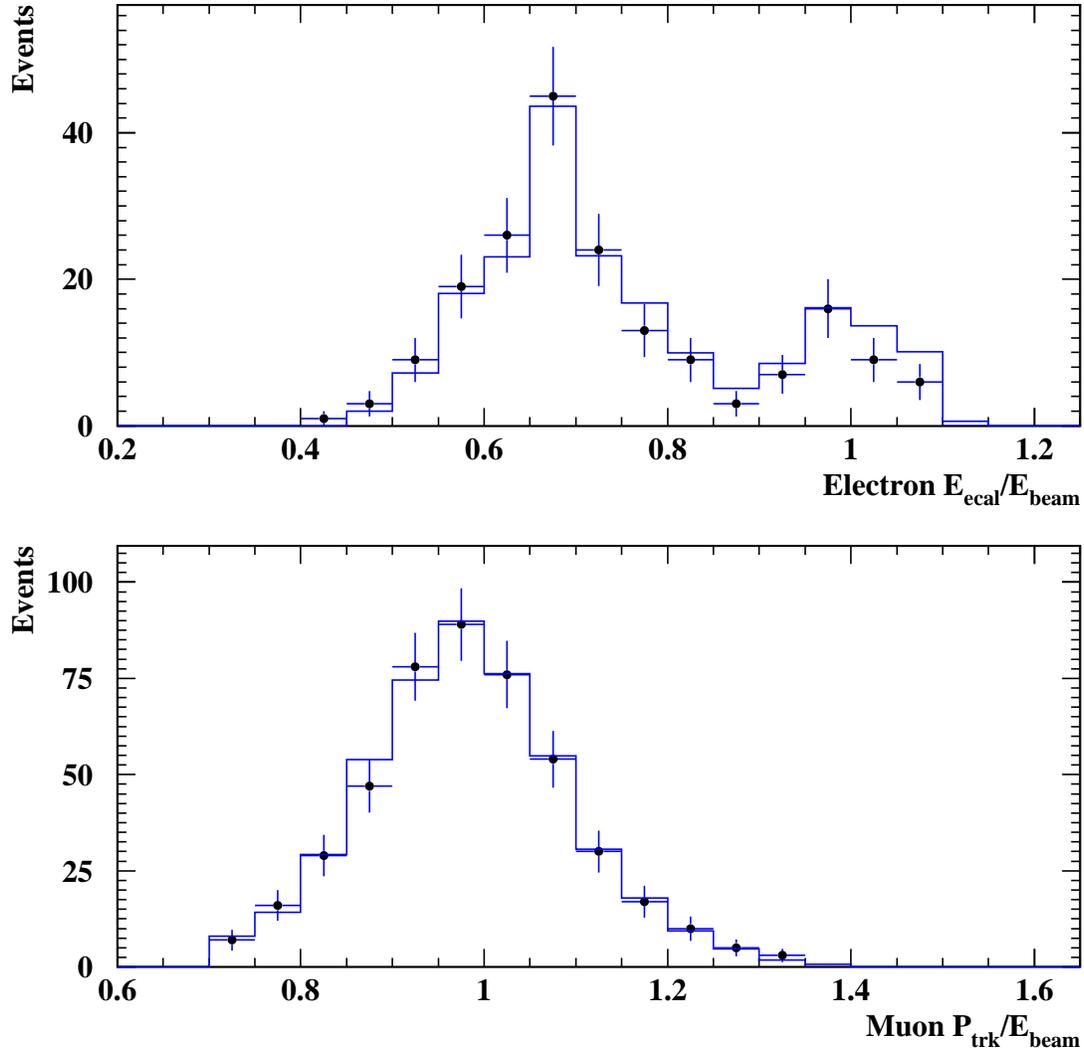}}
       \caption{ The $\rm E_{ECAL}/E_{beam}$  distribution for 
         electron candidates and the $\rm P_{trk}/E_{beam} $ for
         muon candidates  
         for all data (points) as 
         compared to the Standard Model Monte Carlo 
         expectation (solid lines).}   
       \label{fig:plots2}
\end{figure} 
   Events that survive the above cuts are subject to specific \emu, \etau and
   \mutau selection cuts. These cuts are optimised to minimise the
   dependence on $\rm \sqrt{s}$ and to reject SM lepton pair
   final states while keeping reasonable efficiency for  
   \emu, \etau and \mutau.
   \par
\begin{itemize}
\item{ \boldmath \bf Selection for the  \mbox{ $ \rm \bm{e} \bm{\mu} $} channel:}
    The event is required to have only two isolated tracks with opposite 
    charge where each of the tracks belongs to a different jet.
    One of the tracks should be identified as an electron and its 
    measured energy should be within $\rm 15\% $ of the
    beam energy.
    The second track should be identified as a muon candidate with a 
    measured momentum within $\rm 20\% $ of the
    beam energy.
    The total event momentum is required to be less than 
    $\rm 0.25\times E_{beam}. $  In the case  where the opening angle 
    between the two tracks is less than $\rm 170^{\circ}, $ 
    we require in addition that the missing momentum
    should be pointing to the 
    forward part of the detector ($\rm |\cos{\theta}|> 0.9 $). Missing
    momentum caused by undetected neutrinos from a $\rm \tau $ 
    decay would, for the selected events, point
    to the barrel part of the detector.
\item{\boldmath \bf Selection for the  \mbox{$\rm \bm{e}\bm{\tau} $} channel:}
    Here we apply tighter cuts on the electron candidate. 
    The electron should be identified as an isolated track
    with $\rm |E_{ECAL}-E_{beam}|/E_{beam} < 0.10.$ 
    The energy of the recoiling jet against the electron is required
    to be less than $\rm 0.75 \times E_{beam}$, and the total event momentum
    should be larger than 5 GeV and point to the barrel region of the detector.
\item{ \boldmath \bf Selection for the \mbox{$\rm \bm{\mu}\bm{\tau} $} channel:}
     Here we require an identified muon as 
     an isolated track with  $\rm |P_{trk}-E_{beam}|/E_{beam} < 0.15 $.
     The energy of the jet recoiling against the muon is required
     to be less than $\rm 0.75 \times E_{beam}$, and the total event momentum
     should be larger than 5 GeV and point to the barrel 
     region of the detector.
\end{itemize}
     With our lepton identification criteria,    
     a hadron from a single-prong 
     tau decay has a $\rm 3\% $ probability  to be  misidentified  
     as an electron and a $\rm 1.6\% $ probability 
     to be misidentified as muon.  
\begin{table}
   \begin{center}
\caption{Integrated luminosity and 
 efficiency as a function of $\rm \sqrt{s}.$}
 \bigskip
    \begin{tabular}{|c|c|c|c|c|}
       \hline
 $\rm \sqrt{s}(GeV) $ & Lumi. $\rm [pb^{-1}]$ &\emu$ \bm[\%] $ & \etau$\bm[\%]$ & \mutau$\bm[\% ]$ \\
       \hline
       \hline 
 189          & 174.6 & $\rm 56.3\pm 0.7$ & $\rm 24.6\pm 0.5$&$\rm 22.2\pm 0.6$ \\
$\rm 192 < \sqrt{s} < 200 $& 103.6 &$\rm 56.0\pm 0.7$ & $\rm 24.2\pm 0.5$&$\rm 21.8\pm 0.6$ \\
$\rm 200 \leq \sqrt{s} \leq 209  $ & 322.2 &$\rm 55.8\pm 0.7$ & $\rm 23.9\pm 0.5$&$\rm 21.8\pm 0.6$ \\
       \hline       
       \end{tabular}\label{effi}
     \end{center}
\end{table}          
   \begin{table}
   \begin{center}
\caption{ Selected data events versus 
        SM expec\-ta\-tions.
 ``$\rm No-\ell^+\ell^-$'' stands for the cut against events having  
 a pair of electron  ($\rm e^+e^- $) or a pair of muon ($\rm \mu^+\mu^-$)
 candidates in the final state.}
 \bigskip
    \begin{tabular}{|c|c|c|c|c|c|c|}
       \hline
        & \ee & \mumu & \tautau & Other & Total Background & Data \\
       \hline
       \hline 
 Selected                & 20745  & 1359 & 564   & 24    &  22683 & 23164 \\
 $\rm \sum E_{ECAL}< 1.6\ E_{beam} $ 
                            & 2275   & 1359 & 520   & 23    & 4185 &  4201\\
 $\rm No-\ell^+\ell^- $     &  559   & 57   & 67    & 21    & 704  &   713  \\
 \emu  Candidates           &  0     & 0    & 0.015 & 0.004 & 0.019& 1    \\
 \etau Candidates           &  4.010 & 0.017& 0.520 & 0.004 & 5.01 &  5  \\
 \mutau Candidates          &  0.017 & 5.901& 8.400 & 0     &  14.3&  11\\
       \hline       
       \end{tabular}\label{result189}
     \end{center}
\end{table}          
\section{ Results} 
    After the normalisation of the Monte Carlo backgrounds to the 
    integrated luminosity of the data,
    the effects of the selection criteria 
    are  summarised in Table 2.
    The contributions to the SM 
    background from processes other than lepton pair final states
    ($\rm e^+e^-,\mu^+\mu^-\ and\ \tau^+\tau^- $) are very small. These 
    contributions include the full 4-fermion final states, QCD-like 
    final states and $\rm \gamma\gamma $ scattering events. They are
    summed together in the table as a separate column.\par 
     The SM describes the data well for each  $\rm \sqrt{s}$ 
     and each step of the selection procedure, except for a single 
     \emu candidate which was selected at $\rm \sqrt{s} = 189\ GeV. $   
     According to our  Monte Carlo simulations only 0.019 events are 
     expected from SM processes.
     This particular event is displayed in Figure \ref{fig:display1} for 
     the $ r-\phi $ view and in Figure \ref{fig:display2} for 
     the $ r-\theta $ view. 
     It has the following characteristics:
   \begin{itemize}
  \item total visible energy 
       $\rm E_{vis} $ = 176 GeV ($\rm \sqrt{s} =189\ GeV $); 
  \item measured ECAL energy of the electron
      candidate: $\rm E_{ECAL}=(84.83\pm 2.25)\ GeV;$ 
  \item measured momentum of the muon candidate: 
      $\rm (84.5\pm 10)\ GeV$;
  \item opening angle between the two tracks = $\rm 165^o $;
 \item one measured cluster in the forward detector 
       with $\rm E= 7.2\pm 2.5\ GeV $.
  \end{itemize}
    A kinematic fit with energy-momentum conservation is applied 
    assuming that the forward
    detector cluster was due to a photon. The fit 
    has a probability of $\rm 41\%.$\par
    Results of the present search are quantified in terms of 
    $\rm 95\%\ CL$ upper limits of the production 
    cross section of LFV events, assuming uniform 
    final state angular distributions. 
    The upper limits are obtained
    with the method that takes into account the 
    uncertainties on the signal efficiency and 
    on the background expectation as explained 
    in Reference~\cite{stats}. 
    These uncertainties include systematic contributions
    discussed in the next section. 
    The upper limits are displayed in Table 3 for different 
    intervals of $\rm \sqrt{s} $.
    Limits for different assumed angular distributions may be derived
    using the fact that the efficiency is uniform over the barrel region
    of the detector.\par
   \begin{table}
   \begin{center}
\caption{  $\rm  95\%  $ confidence level upper limits on 
 $\rm{\sigma (e^+e^-\give }$ $\rm {e\mu,\ e\tau\ and\ \mu\tau}) $  as a 
 function of $\rm { \sqrt{s} }$.}
 \bigskip
    \begin{tabular}{|c|c|c|c|}
       \hline
Channel & \emu & \etau & \mutau \\
       \hline
$\rm \sqrt{s} (GeV)$ & $\rm \sigma [fb] $& $\rm \sigma [fb] $
 & $\rm \sigma [fb] $ \\
       \hline
       \hline 
 189                              & 58   & 95   & 115   \\ 
 $\rm 192\leq \sqrt{s} \leq 196$  & 62   & 144  & 116   \\ 
$\rm 200 \leq \sqrt{s} \leq 209 $ & 22   & 78   & 64    \\
       \hline       
       \end{tabular}\label{resultetau}
     \end{center}
\end{table}          
\section{Systematic Studies}    
    Five inputs are used to estimate the upper limits: the number of
    selected event candidates, the SM background contribution, 
    the signal selection efficiency and the uncertainties on
    the background expectation and on the signal efficiency.
    These uncertainties are estimated by repeating the analysis while 
    varying the applied cuts, taking into account the following effects:  
\begin{itemize}
\item{\bf Efficiency:}
  The two Monte Carlo methods (see Section 2) yield  compatible results. 
  The data-based method is applied 
  only to the \emu channel, since at each
  $\rm \sqrt{s} $, the reconstructed  \tautau  sample 
  is too small to apply the method.
  A systematic error of $\rm 1\% $ for \emu and of $\rm 2.1\% $ for 
  \etau and \mutau is assigned based on the largest deviation between
  the methods.
\item {\bf Event selection: }
    The following variations in the event selections are made, one at a
    time:
    the maximum number of 
    allowed tracks per event is changed from 8 to 10; 
    the minimum track momentum is 
    increased from 250 to 500 MeV; the cone angle
    is opened from $\rm 15^o\ to\ 30^o; $
    the cone minimum energy is increased from
    $\rm 10\% $ to $\rm 20\%\ of \ E_{beam}; $    
    the cut on the event thrust is set
    to 0.98  instead of 0.95.
\item {\bf Energy and momentum resolution:}
    The electromagnetic calorimeter energy resolution 
    is about $\rm 3.5\% $ for electrons with momenta above 70 GeV.
    The systematic error is estimated 
    by scaling the measured electromagnetic energy of 
    electron  candidates by  $\rm \pm 3.5\%. $ 
    The measured momenta of muon candidates are varied by 
    $\rm \pm 10\% $  to account for the momentum resolution. 
\item{ \bf Integrated luminosity:}
    A  $\rm 0.5\% $ 
    measurement error on the integrated luminosity is added to another 
    $\rm 0.5\% $ interpolation and averaging error between 
    various grouped $\rm \sqrt{s} $ points(see Table 1), 
    after
    which the corresponding error on the SM contribution is calculated.
\end{itemize}  
    The different errors are added in quadrature.  
    The final systematic uncertainty is about $\rm 3.5\% $ on the efficiency 
    and $\rm 5\% $ on the background expectation.
\section{ Conclusion}
    We have no clear evidence for production of  
    lepton flavour violating events such as
    \emu, \etau and \mutau production in 
    \ee collisions between 189 GeV and 209 GeV. 
    We observe  one single \emu candidate, probably  
    produced with initial state radiation, where, 
    according to our Monte Carlo simulations, we expect 0.019 events
    to be produced from Standard 
    Model processes. 
    We have obtained  the first upper limits for
    $\rm \sigma(e^+e^-\give e\mu,\ e\tau\ and\ \mu\tau) $ as a 
    function of $\rm\sqrt{s}$ at LEP2 energies.
    The limits range from 
    22 fb to 58 fb for the \emu channel, from 78 fb to 144 fb for the
    \etau channel and from 64 fb to 166 fb for the \emu channel.\par

\par
\medskip
\bigskip\bigskip\bigskip
\appendix
\par
Acknowledgements:
\par
We particularly wish to thank the SL Division for the efficient operation
of the LEP accelerator at all energies
 and for their continuing close cooperation with
our experimental group.  We thank our colleagues from CEA, DAPNIA/SPP,
CE-Saclay for their efforts over the years on the time-of-flight and trigger
systems which we continue to use.  In addition to the support staff at our own
institutions we are pleased to acknowledge the  \\
Department of Energy, USA, \\
National Science Foundation, USA, \\
Particle Physics and Astronomy Research Council, UK, \\
Natural Sciences and Engineering Research Council, Canada, \\
Israel Science Foundation, administered by the Israel
Academy of Science and Humanities, \\
Minerva Gesellschaft, \\
Benoziyo Center for High Energy Physics,\\
Japanese Ministry of Education, Science and Culture (the
Monbusho) and a grant under the Monbusho International
Science Research Program,\\
Japanese Society for the Promotion of Science (JSPS),\\
German Israeli Bi-national Science Foundation (GIF), \\
Bundesministerium f\"ur Bildung und Forschung, Germany, \\
National Research Council of Canada, \\
Research Corporation, USA,\\
Hungarian Foundation for Scientific Research, OTKA T-029328, 
T023793 and OTKA F-023259.\\
%

\begin{figure}
       \centerline{\includegraphics[scale=0.8]{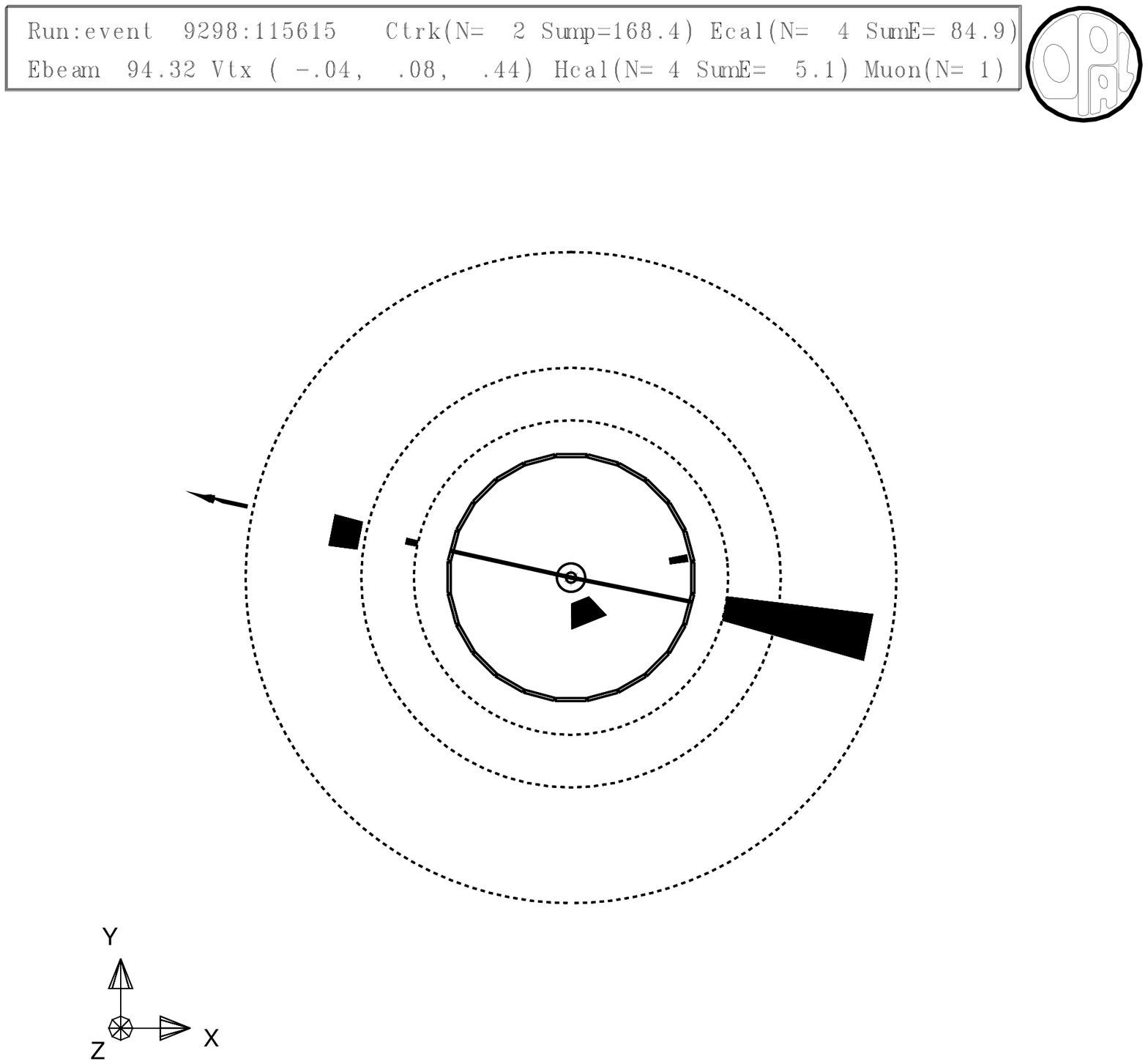}}
       \caption{ An $ r-\phi $ view of the \emu candidate at 189 GeV.
    Two well measured tracks can be seen in the trackers (inner 
    thick circle). The track in the right side deposited all its energy
    in the electromagnetic calorimeter (black trapezoid), and is the electron
    candidate. The track in the left side has a small electromagnetic
    energy deposit and has matching muon hits in the muon chambers (the
    arrow), and is the muon candidate. 
}
       \label{fig:display1}
\end{figure}
\begin{figure}
       \centerline{\includegraphics[scale=0.8]{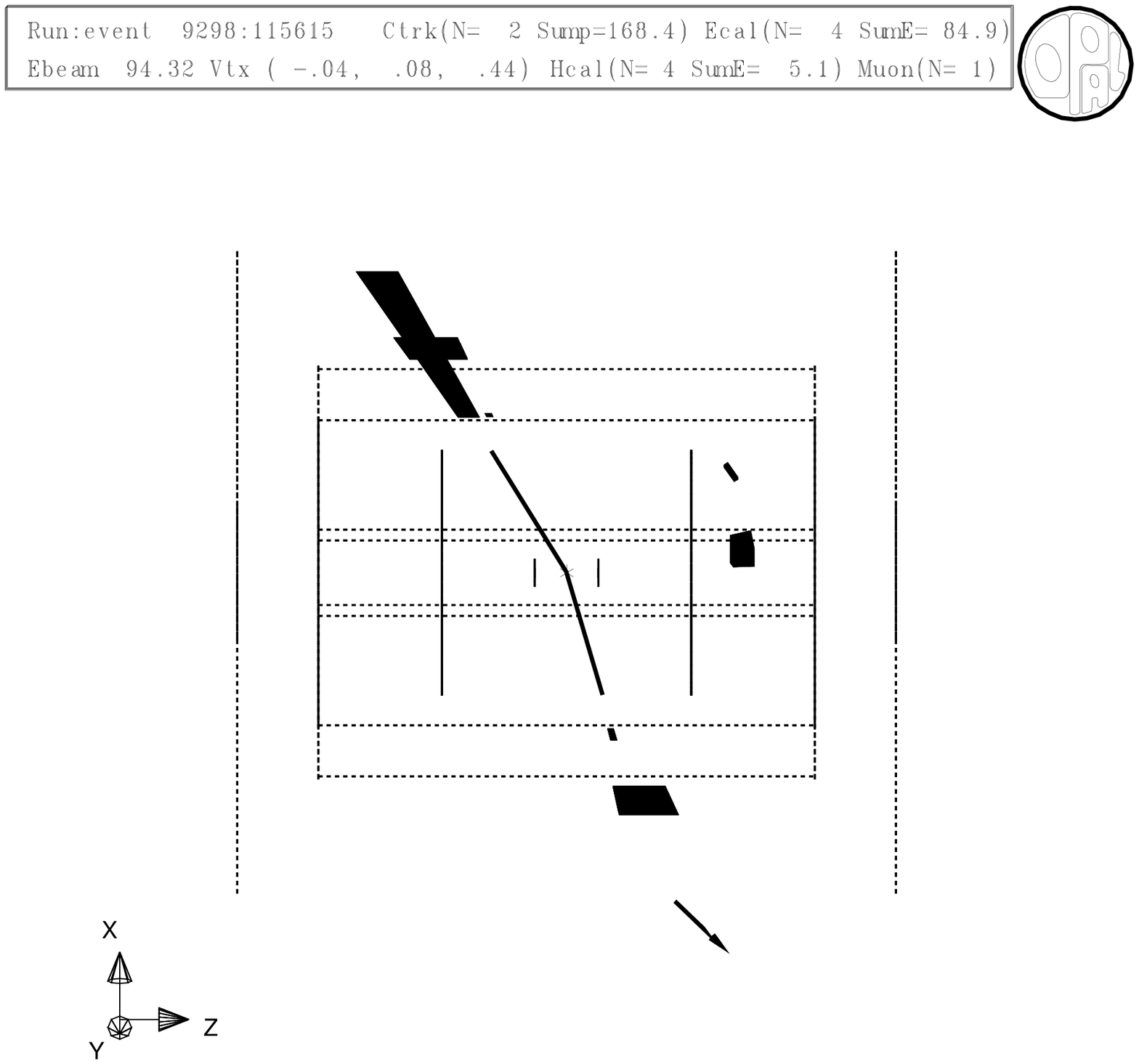}}
       \caption{An $ r-\theta $ view of the \emu candidate at 189 GeV.
     The beam direction is a horizontal line passing trough the 
     intersection of the two tracks.
     The tracks are not back-to-back, and the missing momentum
     is compatible with the observed cluster in the forward detector which
     is the dark block close to the beam axis (right side).}
       \label{fig:display2}
\end{figure} 
\eject
\end{document}